
\magnification=1200

\centerline{\bf Exact Solutions of G-invariant Chiral Equations}
\centerline{\bf by}
\centerline{\bf Tonatiuh Matos}
\vskip.3cm
\centerline{\it Centro de Investigaci\'on y de Estudios Avanzados del I.P.N.}
\centerline{\it Apdo. Postal 14/740, 07000 M\'exico D.F.}
\vskip1.7cm
\centerline {\bf Abstract}

We give a methodology for solving the chiral equations \ \ \ \
$(\alpha g_{,z} g^{-1})_{,\overline z} + (\alpha g_{,\overline z}
g^{-1})_{,z} \ = \ 0 $  \ \ where $g$ belongs to some Lie group $G$. The
solutions are writing in terms of Harmonic maps. The method could be used
even for some infinite Lie groups.

\vskip1cm
One of the most important set of differential equations in mathematical
physics is perhaps the chiral equations. The most popular of them are
the SU(N)-invariant one. In general relativity chiral equations appear
very often for the groups SU(1,1), SU(2,1), SL(2, R) etc. [1]. One of the
first aproaches for solving them is by means of  the Inverse Scatering
Method (ISM), given by Belinsky ans Zacharov [2]. They gave a Lax Pair
respresentation of the chiral equatons and solve them using the ISM. An
equivalent aproach using the polinomial ansatz is given in ref [3]. In this
last work we used the generalized B\"acklund transformations for finding exact
solutions of the chiral equations.

\vskip.3cm
Recently, Hussain \ [4] \ showed that \ self or antiself \ dual gravity
reduces to \
chiral equation-like systems but the group remain here infinite. Algebra
brackets are replaced by Poisson brackets and all of these methods mentioned
above can not be aplied in this case. In this paper we generalize the  method
of Harmonic maps for solving the chiral equations for all Lie groups. All the
solutions obtained by
the ISM or by B\"acklund transformations can be derived with this method
choosing the Harmonic map conveniently. For example, all the solutions found by
ISM in general relativity can also be derived by this mathod [10]. With the
method of Harmonic maps th
e solutions apear clasified in classes naturaly. The solutions can be given in
terms of arbitrary harmonic maps and the method can be used even for infinite
groups.

\vskip.3cm
Let $g$ be a mapping \ $g: {\cal C} \otimes {\cal \overline C} \to G, \ \ \ \
g \to g (z, \overline z) \epsilon G$, where $G$ is a paracompact Lie group.

\vskip.3cm
The chiral equations for $g$

$$ (\alpha g_{,z} g^{-1})_{,\overline z} + (\alpha g_{,\overline z}
g^{-1})_{,z} \ = \ 0 \eqno(1)$$

\noindent
$(\alpha^2 = det \ \ g)$ are a set of non-linear, coupled, second order partial
diffeential equations that appear in many topics in physics. Normally
$g(z, \overline z)$ is given in a representation of $G$. One of the most
important features of the chiral equations is that they can be derived from
the Lagrangian.

$${\cal L} \ = \ \alpha t r \ (g_{,z} g^{-1} g_{,\overline z} g^{-1}).
\eqno(2)$$

This Lagrangian represents a topological quantum field theory with gauge
group $G$, then it is of great interest to know explicit expression of the
elements $g \epsilon G$ in terms of the local coordinates $z$ and $\overline
z$.
\vskip.3cm
Let $G_c$ be a subgroup $G_c \subset G$ such that $c\epsilon G_c$ implies
$c_{,z} = 0,\ c_{,\bar z} = 0$. Then equation (1) is invariant under the left
action $L_c$ of
$G_c$ over $G$. We say that the chiral equations (1) are invariant under this
group.
\vskip.3cm
{\bf Proposition 1.} Let $\beta$ be a complex function defined by

$$ \beta_{,z} \ = \ {1\over 4(ln\alpha)_{,z}} \ tr \ (g_{,z} g^{-1})^2, \ \ \
g\epsilon G \eqno(3)$$

and $\beta_{,\overline z}$ with $\overline z$ in place of $z$. If $g$ fulfills
the chiral equations, then $\beta$ is integrable.

\vskip.3cm
Proof. $\beta_{,z\overline z} \ = \ \beta_{, \overline zz}.$

\vskip.3cm
Let ${\cal G}$ be the corresponding Lie algebra of $G$. The Maurer-Cartan
form $\omega_g$ of $G$ defined by $\omega_g \ = \ L_{g-1*} (g)$ is a one-form
on $G$ with value on ${\cal G}, \omega_g \epsilon T^*_g G \otimes
{\cal G} (T_x M$ represents the tangent space of the Manifold $M$ at the point
$x)$. Let us define the mappings

$$\eqalign{A_z:&G \rightharpoonup {\cal G}\cr& g \rightharpoonup A_z (g) \ = \
g_{,z} g^{-1}\cr}$$

$$\eqalign{A_{\overline z}:&G \rightharpoonup {\cal G}\cr& g \rightharpoonup
A_{\overline z} (g) \ = \ g_{,\overline z} g^{-1}\cr}\eqno(4)$$

If $g$ is given in a representation of $G$, then we can write the one-form
$\omega (g) \ = \ \omega_g$ as

$$ \omega \ = \ A_z dz \ + \ A_{\overline z} d {\overline z} \eqno(5)$$

We can now define a metric on ${\cal G}$ in a standard manner. Since
$\omega_g$ can be written as in (50), the tensor

$$l \ = \ tr (dgg^{-1} \otimes dgg^{-1}) \eqno(6)$$

\noindent
on $G$ defines a metric on the tangent bundle of $G$.

\vskip.3cm
{\bf Theorem 1.} The submanifold of solutions of the chiral equations $S
\subset G$, is a symmetric manifold with metric (6).

\vskip.3cm
This theorem was proved by Neugebauer and Kramer in reference [5] and we will
only outline here the proof.

\vskip.3cm
Proof. We take a parametrization $\lambda^a \ \ a=1, \cdots n,$ of $G$. The
set $\{\lambda^a\}$ is a local coordinate system of the $n$-dimensional
differential manifold $G$. In terms of this parametrization the Maurer-Cartan
one form $\omega$ can be written as

$$ \omega \ = \ A_a d\lambda^a, \eqno(7)$$

where $A_a(g) = ({\partial \over \partial \lambda^a} g) g^{-1}.$ The chiral
equations then read

$$ \nabla_b A_a (g) + \nabla_a A_b (g) = 0, \eqno(8)$$

with $\nabla_a$ the convariant derivative defined by (6).

\vskip.3cm
One infers the relation

$$ \nabla_b A_a (g) = {1\over 2} \ [A_a, A_b] (g). \eqno(9)$$

With this, one can calculate the Riemannian curvature ${\cal R}$ of (6).
Their components read

$$ R_{abcd} = {1\over 4} tr (A_{[a}A_{b]} A_{[c}A_{d]}) \eqno(10)$$

\noindent
$([a,b]$ means index commutation) this can be done, because $G$ is a
paracompact manifold. It follows that $\nabla {\cal R}= 0$

\vskip.3cm
{\bf Proposition 2.} The function $\alpha =$ det $g$ is harmonic.

\vskip.3cm
Proof. The trace of the chiral equations implies $\alpha_{,z\overline z} =
 0.$

\vskip.3cm
In the following we will explain the method for calculating explicit chiral
field.

\vskip.3cm
Let $V_p$ be a complete totally geodesic submanifold of  $G$ and let $\{
\lambda^i \} \ \ \ i = 1, \cdots , p$ be a set of local coordiantes on $V_p$.
Because we know the Manifold $G$, it is possible to know $V_p$. In fact
$V_p$ is a subgroup of $G$ and since $V_p$ is a complete totally geodesic
submanifold, $V_p$ is also symmetric. The symmetries of $G$ and $V_p$ are
in fact isometries, since both of them are paracompact manifolds, with
Riemannian
metrics (6) and $i_*l$ respectivelly. ($i$ is the restriction of $V_p$ into
$G$). Let us suppose that $V_p$ possesses $d$ isometies. Thus

$$ (\alpha \lambda^i_{,z})_{,\overline z} + (\alpha \lambda^i_{,\overline
z})_{,z} + 2 \alpha \sum _{ijk} \Gamma^i_{jk} \lambda^j_{,z} \lambda^k_
{,\overline z} \ = \ 0 \eqno(11)$$

$$ i, j, k = 1, \cdots  ,p $$

\noindent
where $\Gamma^i_{jk}$ are the Christoffel symboles of $i_*l$ and $\lambda^i$
are the totally geodesic parameters on $V_p$. In terms of the parameters
$\lambda^i$ the chiral equations read

$$ \nabla_i A_j (g) + \nabla_j A_i (g) = 0 \eqno(12)$$

\noindent
where $\nabla_i$ is the covariant derivative of $V_p$. Equation (12) is the
Killing equation on $V_p$ for the components of $A_i$. Since we know the
manifold $V_p$, we know its isometries and therefore its Killing vector
space. Let $\xi_s,  s=1,\cdots , d, $ be a base of the Killing vector space
of $V_p$ and $\Gamma^s$ be a base of the subalgebra corresponding to $V_p$.
Then we can write

$$ A_i (g) = \sum _s \ \xi^i_s \ \Gamma^s \eqno(13)$$

\noindent
where $\xi_s \ = \ \sum _j \ \xi^j_s \ {\partial \over \partial \lambda^j}.$
The
covariant derivative on $V_p$ is given by

$$ \nabla_j A_i (g) = - {1\over 2} [A_{i,} A_j] (g) \eqno(14)$$

\noindent
where $A_i$ fulfills the integrability conditions

$$ F_{ij} = \nabla_j A_i (g) - \nabla_i A_j (g) - [A_{j,}A_i](g) = 0
\eqno(15)$$

\noindent
i.e., $A_i$ has a pure gauge form.

\vskip.3cm
The left action of $G_c$ over $G, L: G_c \times G \rightharpoonup G$ must
be defined in a convenient manner in order to preserve the properties of the
elements of $S$.

\vskip.3cm
Knowing $\{\xi_s\}$ and $\{\Gamma^s\}$ one could integrate the elements of
$S$, since $A_i (g) \epsilon {\cal G}$ can be mapped into the group by means
of the exponential map. Nevertheless it is not possible to map all the
elements one by one. Fortunately we have the following proposition.

\vskip.3cm
{\bf Proposition 3.} The relation $A^c_i r A_i$ iff there exist $c\epsilon
G_c$ such that $A^c = A \circ L_c$, is an equivalence relation.

\vskip.3cm
Proof. Trivial.

\vskip.3cm
This equivalence relation separates the set $\{A_i\}$ into equivalence
classes $[A_i]$. Let $TB$ be a set of representatives of each class,
$TB = \{[A_i]\}$. Now we map the elements of $TB \subset {\cal G}$ into the
group $S$ by means of the exponential map or by integration. Let us define
$B$ as the set of elements of the group, mapped from each representative
$B = \{g \epsilon S|g = exp (A_i), A_i \epsilon TB\}\subset G.$ The elements
of $B$ are also elements of $S$ because $A_i$ fulfills the chiral equations,
$i.e. B \subset S$. For constructing all the set $S$ we have the following
Theorem.

\vskip.3cm
{\bf Theorem 2.} $(S, B, \pi , G_c, L)$ is a principal fibre bundle with
projection $\pi(L_c(g)) = g; \ \ \ L(c,g) = L_c (g)$.

\vskip.3cm
Proof. The fibres of $G$ are the orbits of the group $G_c$ on $G,F_g =
\{g^\prime \epsilon G|g^\prime = L_c (g)\}$ for some $g\epsilon B$. The
topology of $B$ is its relative topology with respect to $G$. Let $\alpha_F$
be the bundle $\alpha_F = (G_c \times U_{\alpha , } U_{\alpha}, \pi)$, where
$\{U_\alpha\}$ is an open covering of $B$. We have the following lemma.

\vskip.3cm
{\bf Lemma 1.} The bundle $\alpha_F$ and $\alpha = (\pi^{-1} (U_\alpha),
U_{\alpha ,}\pi|_{\pi^{-1}(U_a)})$ are isomorphic.

\vskip.3cm
Proof. The mapping

$$ \psi_\alpha: \phi^{-1} (U_\alpha) = \{g\epsilon S|g^\prime = L_c (g),
g \epsilon U_\alpha\}_{c\epsilon G_c} \rightharpoonup G_c \times U_\alpha$$

$$ g^\prime \rightharpoonup \psi_\alpha (g^\prime) = (c,g)$$

\noindent
is an homeomorphism and $\pi|_{\pi^{-1}(U_{\alpha})} (g^\prime) = g = \pi_2
\circ \psi_\alpha (g^\prime).$

\vskip.3cm
By  lemma 1 the bundle $\alpha$ is locally trivial. To end the proof of the
Theorem it is sufficient to prove that the $G_c$ spaces $(S,G_c, L)$ and
$(G_c \times U_{\alpha ,} G_{c,}\delta)$, are isomorphic,but that is so
because

$$ \delta \circ id|_{G_c} \times \psi_\alpha = \psi_\alpha\circ L|_{G_c
\times \pi^{-1}(U_\alpha)}$$

With this theorem it is now possible to explain the method.

\vskip.3cm
a) Given the chiral equations (1), invariant under the group $G$, chose a
symmetric Riemannian space $V_p$ with a $d$ dimensional isometry group
$H\subset G$, $p \leq n =$ dim
$G$.

\vskip.3cm
b) Look for a representation for the corresponding Lie Algebra ${\cal G}$
compatible with the commutating relations of the killing vectors, via
equation (14).

\vskip.3cm
c) Write the matrices $A_i(g)$ explicitly in terms of the geodesic parameters
of the symmetric space $V_p$.

\vskip.3cm
d) Use proposition 2 for finding the equivalence classes in $\{A_i\}$ and
choose a set of representatives.

\vskip.3cm
e) Map the lie algebra representatives into the group.

\vskip.3cm
The solutions can be constructed by means of the left action of the $G_c$
group into $G$.

\vskip.3cm
Let us give an example. Suppose $G = SL (2, {\bf R}).$

\vskip.3cm
a) We choose the one dimensional space $V_1$ with $i_*l = d \lambda^2$.  It
is Riemannian and symmetric with one killing vector

\vskip.3cm
b) The algebra $sl (2, {\bf R})$  is the set of traceless and real $2 \times 2$
matrices. The Killing equation reduces to the Laplace equation $(\alpha
\lambda_{,z})_{,\overline z} + (\alpha \lambda_{,\overline z})_{,z} = 0.$

\vskip.3cm
c) Using (14) we obtain

$$ g_{,\lambda} g^{-1} = A = constant. $$

\vskip.3cm
d) The representative of the set of trasless and real constant $2\times 2$
matrices $\{A_i\}$ is

$$\left\{[A_i]\right\}=\left\{\left(
\matrix{0 &1\cr
        a &0\cr }\right)\right\} $$

\vskip.3cm
\noindent
(see also ref. [6]).

\vskip.3cm
e) The mapping of the Lie algebra representatives can be done by integration of
the matrix differential equation

$$ g_{,\lambda} = [A_i] g.$$

\vskip.3cm
The solutions depend on the characteristical polynomial of $[A_i]$.
We obtain three cases: $a>0, \ \ a<0, \ \ a=0.$ For each case the matrix
$g$ reads

$$ \eqalignno{
a&>0 \ \ \ g = b \pmatrix{1&a\cr a&a^2\cr} e^{a \lambda} + c \pmatrix{1&-a\cr
-a&a^2\cr} e^{-a\lambda}, \ 4bca^2=1 \cr
a&<0 \ \ \ g = b \pmatrix{cos(a\lambda +\psi_0) & -a sin (a \lambda + \psi_0)
\cr -a sin (a \lambda + \psi_0) & -a^2 cos (a \lambda + \psi_0)\cr}, a^2 b^2
= -1\cr
a&=0 \ \ \ g = \pmatrix{b \lambda + c & b \cr b & 0 \cr} \ \ \ \ \ \ \ \ \
\ \ \ b^2 = -1 \cr}$$

\noindent
where $a, b, c$ and $\psi_0$ are constants. So, for each solution
$\lambda$ of the Harmonic equation $(\alpha \lambda_{,\bar z})_{,z}+
(\alpha \lambda_{, z})_{,\bar z}= 0$ we will
have a new solution of the chiral one. The left action of $SL_c (2, R)$ over
$ SL (2, R)$ is represented as

$$ g^\prime \ = \ C g D$$

\noindent
where $C, D \epsilon S L_c (2, R)$. $g^\prime$ will be also a solution of the
chiral equations.

\vskip.3cm
If we choose a $V_2$ manifold we can fiend an other class of solutions. All
$V_2$ manifold is conformally flat, therefore the metric on $V_2$ reads

$$ d l^2 \ = \ {d \tau d \lambda \over (l + k \lambda z)^2}$$

But $V_2$ symmetric implies $k =$constant. A $V_2$ manifold with constant
curvature has three Killing vectors. Let

\vskip.3cm

$$\eqalignno{ \xi_1 &= {1\over 2V^2} \ [(k\tau^2 + 1) d \lambda + (k \lambda
+ 1) d \tau]\cr \xi_2 &= {1\over V^2} [ - \tau d \lambda + \lambda d \tau] \
\ \ \ \ \ \ \ \ V = 1 + k \lambda \tau \cr \xi_3 &= {1\over 2V^2} [(k\tau^2
- 1) d \lambda + (1 - k \lambda^2) d \tau ]\cr}$$

\noindent
be a base of the Killing vector space of $V_2$. The correspondings
commutation relations of the dual vectors are

$$\eqalignno{[\Gamma^1, \Gamma^2] &= -4 k \Gamma^3 &(16) \cr [\Gamma^2,
\Gamma^3]
 &= 4 k \Gamma^2 \cr [\Gamma^3, \Gamma^1] &= -4 \Gamma^2 \cr}$$

 We have to put $k = -1$ in order to have the commutation relations of
 $sl (2, R)$. A representation of $sl (2,R)$, compatible with (16) is

 $$ \Gamma^1 = 2 \ \pmatrix{ -1&0 \cr 0 & 1\cr}, \ \ \Gamma^2 = \pmatrix{ 0&b
 \cr a & 0\cr}, \ \ \Gamma^3 = \pmatrix{0 & -b \cr a & 0 \cr} \ \ \ ab =1$$

 Then equation (13) reads

 $$\eqalignno{ g_{,\lambda} g^{-1} &= A_\lambda = {1\over V^2}
 \pmatrix{\tau^2 -1 & b(1-\tau)^2 \cr -a (1+\tau)^2 & 1-\tau^2\cr} \cr
g_{,\tau}
 g^{-1}&= A_\tau = {1\over V^2} \pmatrix{\lambda^2 -1 & b(1-\lambda)^2 \cr
 -a (1+\lambda)^2 & 1-\lambda^2 \cr} \cr}$$

\noindent
wich integration is

 $$ g = {1\over 1 - \lambda \tau} \ \ \pmatrix{ c (1-\lambda)(1-\tau) & e(\tau
 - \lambda ) \cr e (\tau -\lambda) & d (1+\lambda)(1+\tau)\cr}$$
$$cd =-e^2,\ \ \ a={e\over c},\ \ \ b=-{e\over d}  \eqno(17) $$

The harmonic equation on $V_2$ reads

$$ \eqalignno{ (\alpha \lambda_{,z})_{,\overline z} &+ (\alpha \lambda_
{,\overline z})_{,z} + {4\tau \over 1+\lambda \tau} \alpha \lambda_{,z}
\lambda_{,\overline z} = 0 \cr (\alpha \tau_{,z})_{,\overline z} &+ (\alpha
\tau_{,\overline z})_{,z} + {4 \lambda \over 1 + \lambda \tau} \alpha
\tau_{,z} \tau_{,\overline z} = 0&(18)\cr}$$

So, for each solution of (18) into (17) we fiend a new solution of the chiral
equations for  $SL (2,R)$.

\vskip.3cm

A general algorithm of integration even for the Lie Algebra $sl(N,{\bf R})$, is
given
in [7]. The explicity example for $SL(4, R)$ is presented in ref. [8].
The application of the method for infinite groups will be published elsewhere
[9].

\vskip.3cm
{\bf Acknowledgments.}
\vskip.3cm
This work is supported in part by CONACyT-M\'exico.

\vskip.5cm
{\bf References.}
\vskip.3cm
\item{[1]} Matos T. Chiral Equation in Gravitational Theories. In Proceedings
of the
International Conference on Aspects of General Relativity and Mathematical
Physics. Ed. by N. Bret\'on, R. Capovilla and T. Matos. CINVESTAV 1994.
\vskip.3cm
\item{[2]} Belinsky V.A. and Zacharov V.E.{\it Zh. Eksp. Teor. F\'\i s}, {\bf
15},
(1978), 1953.
\vskip.3cm
\item{[3]} Kramer D., Neugebauer G. and Matos T. {\it J. Math. Phys}, {\bf 32},
(1991),
2727.
\vskip.3cm
\item{[4]} Hussain V. {\it Phys. Rev. Lett}, {\bf 72}, (1994), 800.
\vskip.3cm
\item{[5]} Neugebauer G. and Kramer D. B\"acklund Transformations of
Einstein-Maxwell Fields. \ \ Preprint, Friederich-Schiller-Universit\"at
Jena, (1991), Germany.
\vskip.3cm
\item{[6]} Matos T. {\it Ann. de Phys. (Leipzig)}, {\bf 46}, (1989) 462.
\vskip.3cm
\item{[7]} Matos T., Rodr\'\i guez L. and Becerril R. {\it J. Math. Phys}. {\bf
33},
(1992), 3521.
\vskip.3cm
\item{[8]} Matos T. and Wiederhold P. {\it Lett. Math. Phys}. {\bf 27} (1993),
265.
\vskip.3cm
\item{[9]} Matos T., Garc\'\i a-Compe\'an, H. and Capovilla, R. In preparation.
\vskip.3cm
\item{[10]} Matos T., Plebanski J. {\it Gen. Rel. Grav.} {\bf 26}, (1994).

\end